\documentstyle[preprint, aps, prl]{revtex}

\begin{document}
\preprint{Preprint: LANL (1999)}

\title 
{\Large
{Structure and Dynamics of Surface \\
Adsorbed Clusters}
}

\author{J.R.Sanchez}

\address
{
Facultad de Ingenier\'{\i}a \\
Universidad Nacional de  Mar del Plata \\
Av. J.B. Justo 4302 \\
7600 Mar del Plata \\
Argentina \\
}

\vspace {1 truecm}


\maketitle
 
\begin{abstract}
\mediumtext
 
Extensive numerical simulation are reported for the structure and
dynamics of large clusters on metal(100) surfaces. 
Different types of perimeter hopping processes makes center-of-mass 
of the cluster to follow a a random walk trajectory. 
Then, a {\it diffusion coefficient} $D$ can be defined as 
$\lim\limits_{t\rightarrow \infty} D(t)$, with $D(t)=\langle d^2 \rangle/(4t)$ and 
$d$ the displacement of the center-of-mass.  
In the simulations, the dependence of the diffusion coefficient on those perimeter 
hopping processes can be analyzed in detail, since the relations between different rates 
for the processes are explicitly considered as parameters.  
\end{abstract} 
 
\newpage 

The different diffusion processes that take place on surfaces have, clearly, 
an important role in many technological areas.
Diffusion of individual atoms and clusters has been studied for a long time with
different experimental techniques and, more recently, using scanning tunneling
microscopy (STM)~\cite{SOL2,SOL3}.
From a theoretical point of view, several lattice-gas kinetic Monte Carlo
simulations~\cite{BIN} have addressed the question of the dependence of the
cluster diffusion coefficient$D$ on the cluster size in square~\cite{SOL4,SOL5}
or triangular~\cite{SOL6} lattice.
Recently, the diffusion of metal clusters on metal surfaces has received 
extensive attention due to the fact that some
experimental and theoretical studies have led to the expectation 
that only small, two-dimensional (2D) clusters were able to diffuse. 
The larger two-dimensional clusters, also observed on the surfaces, 
were not expected to diffuse.
However, recent experimental evidence from STM studies~\cite{EXP} have became 
available showing that very large 2D Ag clusters clusters 
(containing $N = 10^2 - 10^3$ atoms) are indeed able to diffuse 
on Ag(100) substrates.

From the previous theoretical and experimental work, it is now clear that
the movement of the cluster is a consequence of several 
atomic-scale processes, taking place at the periphery of the cluster, 
that makes the center-of-mass follow a random walk trajectory.  
For instance, for small clusters ($N < 20$) it has been proposed that the 
mechanism of diffusion is the short range motion of a single atom away from the 
periphery followed by a regrouping of the cluster around the 
peripheric vacancy~\cite{WANG6}. 
Evidence of concerted gliding is also available~\cite{FINK5}.
On the order hand, the movement of clusters composed of a large 
amount of atoms has been considered a consequence of two main 
mechanisms taking place at the boundary of the cluster: periphery 
diffusion (PD) and 2D evaporation-condensation (EC).
In the PD process several types of atomic motions {\it along} 
the periphery of the cluster are responsible for the displacement of 
the center-of-mass. 
However, the atoms do not leave the cluster while executing
those movements.
In the EC process the cluster is considered to be in quasi-equilibrium 
with a dilute 2D gas of atoms, diffusing very quickly on the metal 
surface surrounding the cluster.
Both types of process are non-exclusive and can take place at the same time.
Of course, it can be anticipated that the energy barriers for the PD mechanism are
much lower than the energy needed for an atom to leave the cluster and, in principle,
the PD process can be expected to be primarily responsible for the movement of 
the cluster.

At this point, the question of the dependence of the diffusion coefficient 
on the number of atoms in the cluster ($N$) becomes relevant.
The diffusion coefficient $D$ can be defined as
\begin{equation}  
\lim\limits_{t\rightarrow \infty} D(t) =  \langle d^2 \rangle/(4t) \:,
\end{equation}
with $d$ the displacement of the center-of-mass.

It has been suggested that the value of the diffusion coefficient $D$ behaves 
as $D \sim N^{-\alpha}$. 
Different values for the exponent $\alpha$ 
have been proposed depending on which diffusion mechanism is considered 
to facilitate the movement of the cluster \cite{SOL1,SOL2}. 
In this sense, Monte Carlo simulations for cluster diffusion
based on the PD mechanism are available, showing 
$\alpha \cong 1.5$~to~$2.0$~\cite{KANG14,VOTER18}.
However, these values of $\alpha$ lead to a strong variation of the diffusion
coefficient $D$ as a function of the number $N$, which is not consistent with 
the experimental STM data available~\cite{EXP}.
On the other hand, in reference~\cite{VANSIC} it is claimed that for 
{\it circular} clusters the diffusion coefficient behaves as
$D \sim d^{-n}$, where $d$ is the {\it diameter} of the cluster and the
integer $n$ identifies different diffusion mechanisms. 
When the center of mass motion occurs by adatom diffusion along the 
periphery of the cluster $n = 3$, while $n = 2$ or $1$ when the 
cluster diffusion occurs by correlated or uncorrelated adatom evaporation and
condensation, respectively. 

Then it is clear that a deeper understanding of the way in which 
2D large clusters move can be obtained using numerical simulations.
Following this idea, we set up Monte Carlo simulations for the kinetics 
in which each of the above described processes takes place at its own rate. 
Diffusion of atoms along the cluster perimeter proceeds at a rate $r_e$ 
and evaporation-condensation at a rate $r_c$. 
Both are related by the detailed-balance relation  
${\left({r_c / r_e}\right)} = exp(-\Delta E/k_bT)$, 
where $\Delta E$ is the energy difference between the two processes. 
The relation between the rates is kept as a parameter in the simulations. 

Our Monte Carlo simulation proceeds as follows. We start with a  
`square' cluster of $N$ atoms at the center of a $1500 \times 1500$ lattice  
substrate~\cite{SQR}.
At a randomly selected site from the periphery of the cluster two main events can 
take place. 

(a) With probability $r_c/r_e$, an evaporation or condensation event is selected.
If the evaporation event is going to take place, an atom is detached from the 
cluster and goes into the surrounding two dimensional gas of 
fast, diffusing atoms. 
If the condensation event is going to take place, an atom is attached 
to a randomly selected place on the periphery of the cluster since it 
is considered to come from the surrounding 
two dimensional gas.
In this way, the average number of atoms in the cluster in maintained equal to $N$.

(b) With probability $1-(r_c/r_e)$, an atom is relaxed to the nearest-neighbor or  
next-nearest-neighbor empty site where it finds that more bonds are saturated. 
This is the mechanism that makes the cluster stay connected 
(i.e. not dissolve) in the $r_e \gg r_c$ limit.
We restrict ourselves within this limit letting 
$r_c/r_e$ be less than 0.2.  

After a certain thermalization time, we start to record the position of 
the center-of-mass of the cluster with respect to the origin of coordinates considered to 
be at the center of the lattice representing the substrate.
Time is measured in Monte Carlo steps per atom. 

The simulational results are shown in Figs 1 to 3. 
Results presented have been averaged over 100 independent runs and for 10 
different cluster sizes containing from 121 to 961 particles.

In Fig. 1 is a representative plot of the overall behavior of the 
diffusion coefficient. 
It can be seen that, in agreement with 
the available experimental results~\cite{EXP},
the simulations show a fast decay in the value of the diffusion coefficient 
at early times, for 
${\left({r_c / r_e}\right)} = 0.1$ and $121 \leq N \leq 961$. 
After this fast decay the cluster reaches a diffusive random-walk-like behavior, 
demonstrated by the plateau at $t \rightarrow \infty$.
This figure confirms that the  main characteristics of the
movements of the cluster are well reproduced in the simulations.
In particular, the very fast decay at early times has been associated with a
certain back correlation effect, and it has been observed in previous cluster 
diffusion simulations where it was considered to be related to the 
cluster connectivity~\cite{KANG14}.

In Fig. 2 the diffusion coefficient $D$ is plotted versus $1/N$. 
From the plot it can be seen that $D$  depends on $N$ as 
$D = D_0{\left({r_c / r_e}\right)} N^{-\alpha}$ with $\alpha = 1$. 
The most notable characteristic
is the dependence of the pre-factor $D_0$ on the evaporation-relaxation relation
$r_c / r_e$. 
The dependence can only be observed by a numerical simulation
that allows full control of the relation. 
It can be seen that the cluster has a greater
$D$ for greater values of the relation $r_c / r_e$. 
This result is a little bit surprising. 
It tell us that, although the predominant processes at the periphery is the PD process, 
the EC process has a greater effect on the {\it mobility} of the cluster. 
Then, the simulations favor the EC mechanism as the primary one 
responsible for the movement of the cluster, in accordance with 
the available experimental data.

From Fig. 3, it can be seen that the exponent $\alpha$ does not depend
on the ratio $r_c / r_e$. 
By linear fitting the points of the plot, an average~\cite{AVR}
value for the exponent $\alpha$ is obtained. 
This value is $\alpha = 1.092 \pm 0.123$ and differs from the value 
obtained from the experimental data, just reported in reference~\cite{EXP}.
However, the experimental measurements of $D$ as a function on $N$ 
have been adjusted within a certain experimental error 
and cannot be considered to be conclusive.
On the other hand the value of $\alpha \sim 1.0$ is in accordance with the
theoretical predicted value of reference~\cite{VANSIC} ($n = 1$) when the
evaporation-condensation process is uncorrelated.

Finally, in Fig 4 the behavior of the diffusion coefficient vs
time (at early evolution times) is plotted in log-log plot for $r_e = 1.0$ and
three different values of $r_c$. 
From this plot a very new characteristic of the movement is identified.
The diffusion coefficient behaves as $D(t) \sim t^{-\beta}$ 
with $\beta = 0.812 \pm 0.021$, showing a nontrivial scaling behavior 
at early times.
In principle, it could be anticipated that the scaling behavior should 
be in close relation with the above mentioned back correlation effect. 
But, the explanation of this characteristic from a physical point of view
needs more experimental and theoretical work.

In conclusion.
Results for the simulation of the diffusion of large Ag clusters over Ag(100) surfaces 
are presented.
The simulations reproduce very well the main characteristic of the movement
of the cluster according to the available experimental results.
The dependence of the diffusion coefficient on the number of atoms in the
cluster has been explored, as well as other features of the mechanisms by which the
large clusters diffuses. 
Also, from the simulations new characteristics of the movement have been identified
and suggest new experimental and theoretical work.
A deeper understanding of the basic cluster diffusion mechanisms 
will have important consequences on the techniques that make possible the 
epitaxial growth.



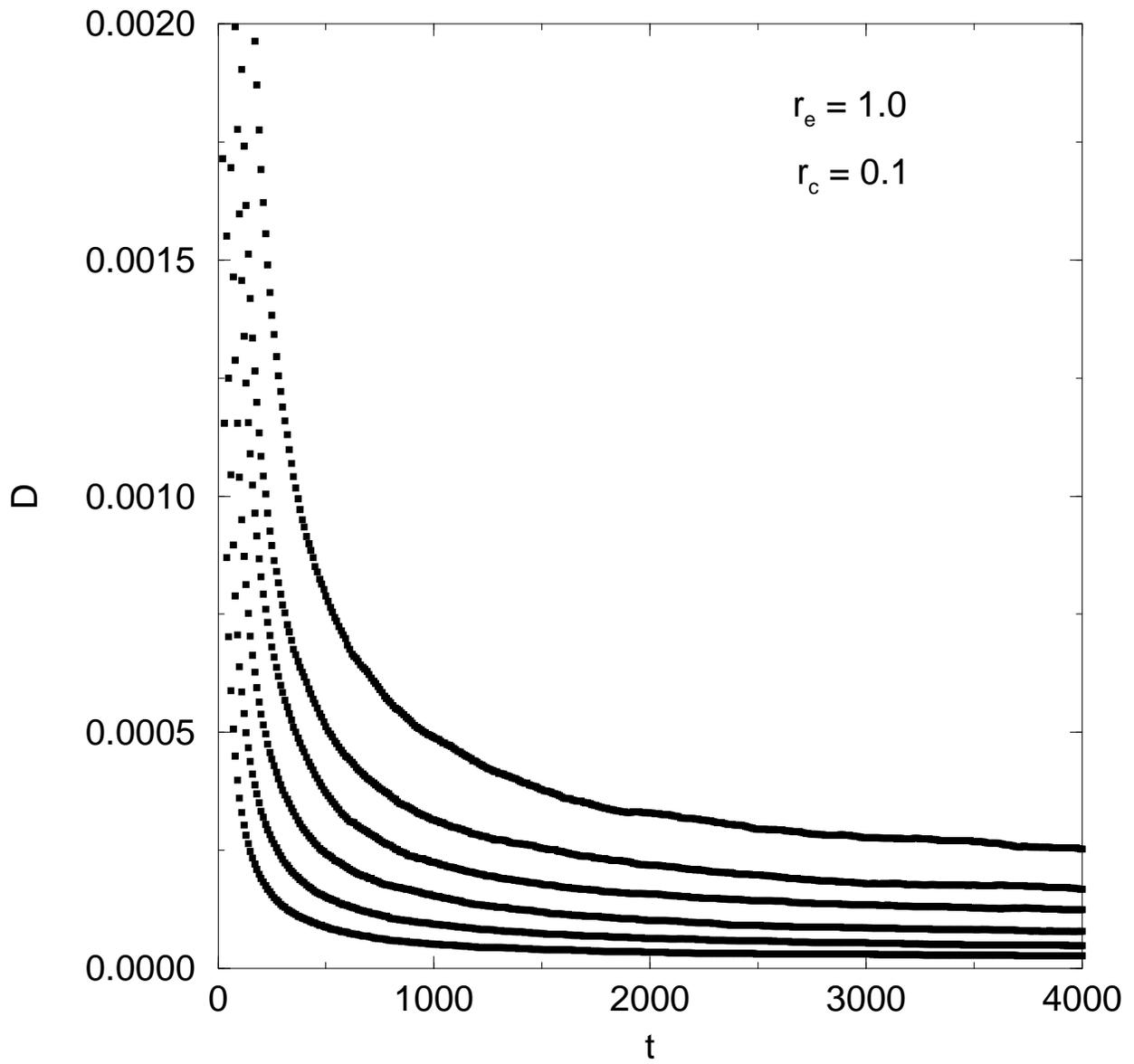
\begin{figure}
\caption{The behavior of the diffusion coefficient along the total simulation
time for $N = 121$ (top curve) to  $N = 961$ (bottom curve). 
The very fast decay at early time is in 
correspondence with the experimental results.}
\label{fig1}
\end{figure}

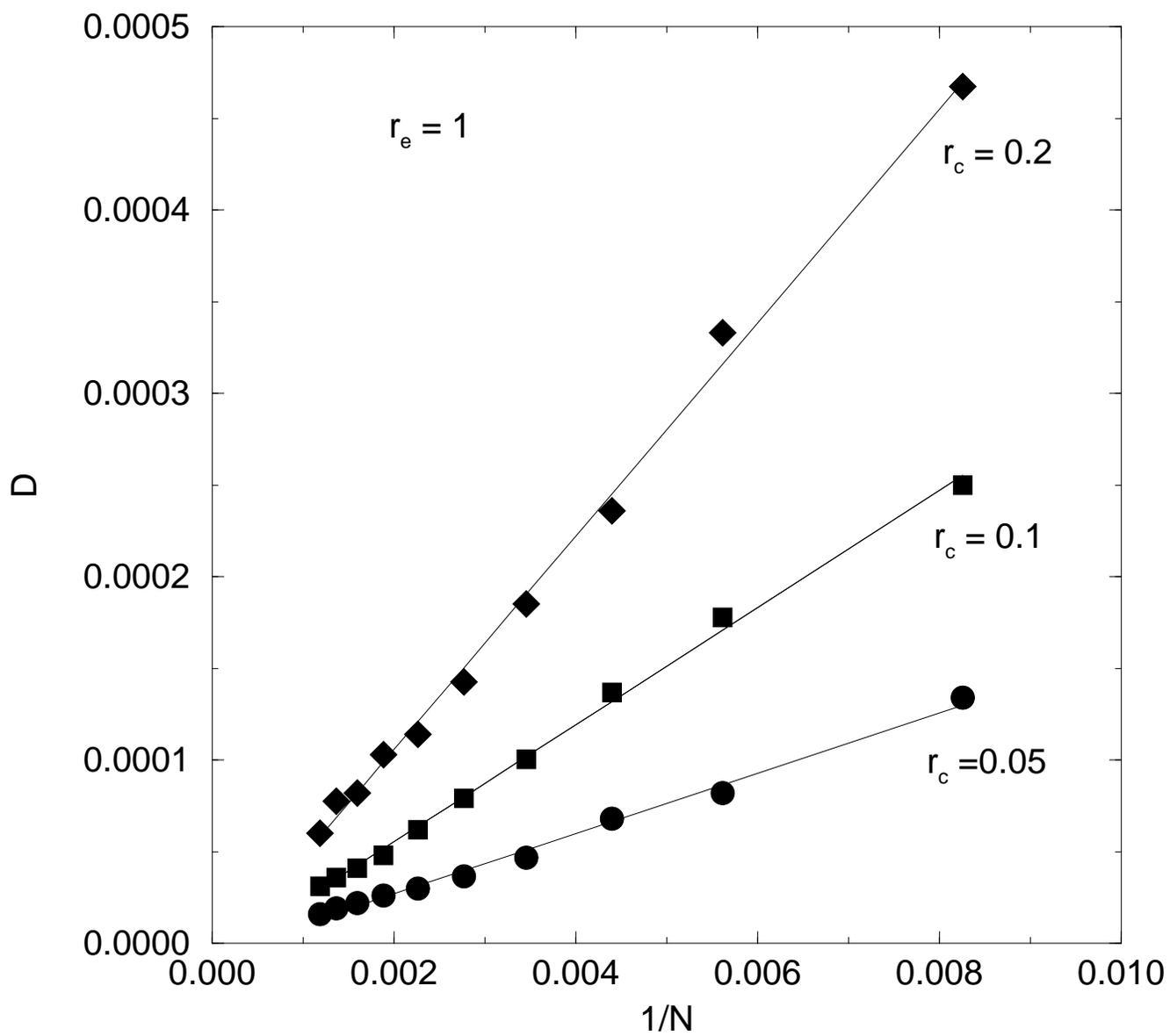
\begin{figure}
\caption{The diffusion coefficient $D$ versus $1/N$. Different slopes for different
values of $r_c$ indicate that the prefactor $D_0$ depends on the $r_c / r_e$ relation.}
\end{figure}

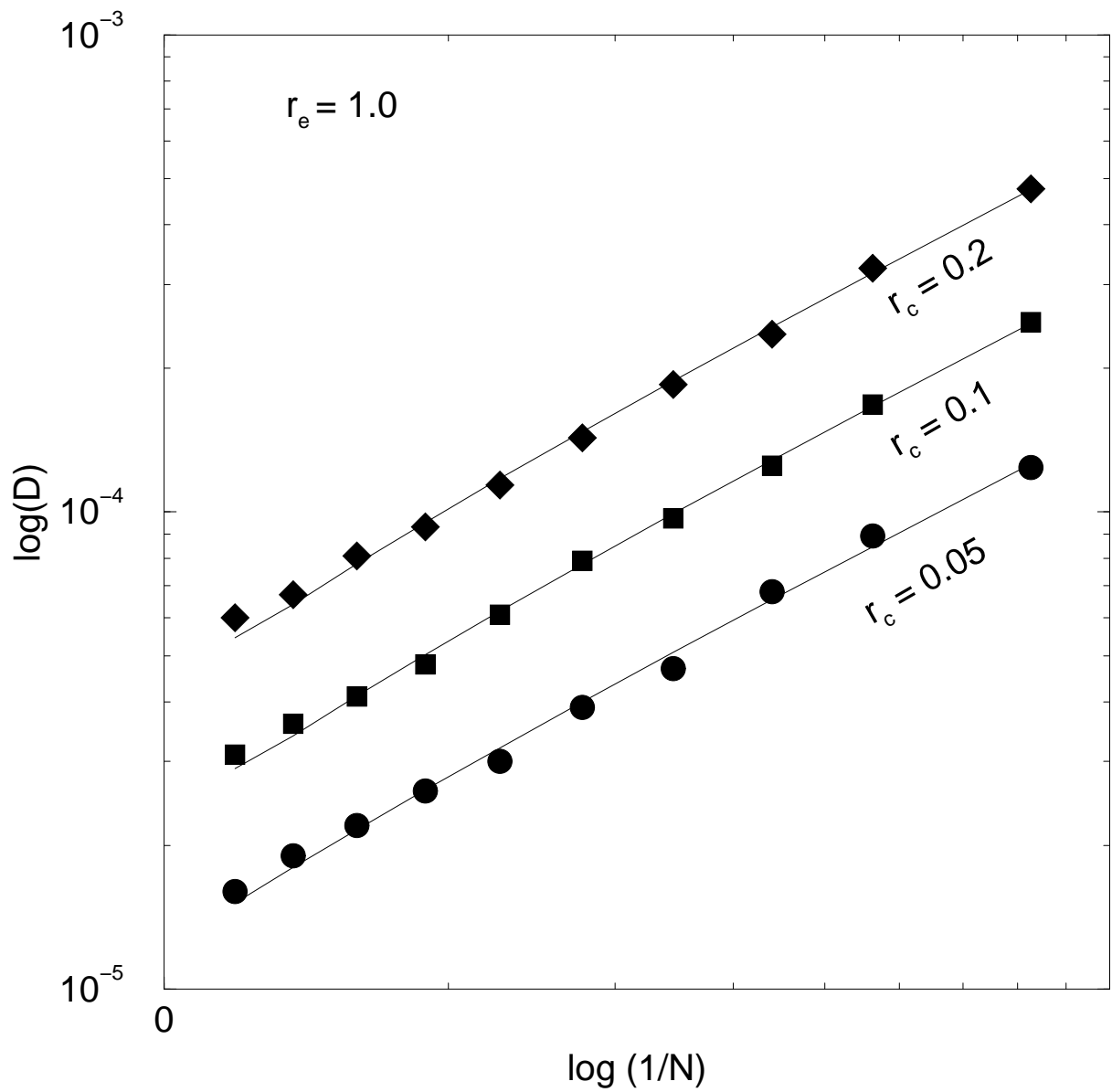
\begin{figure}
\caption{Log-log plot of $D$ versus $1/N$. The average slope of the three plots gives  
the value for the exponent $\alpha = 1.092 \pm 0.123$.}
\end{figure}

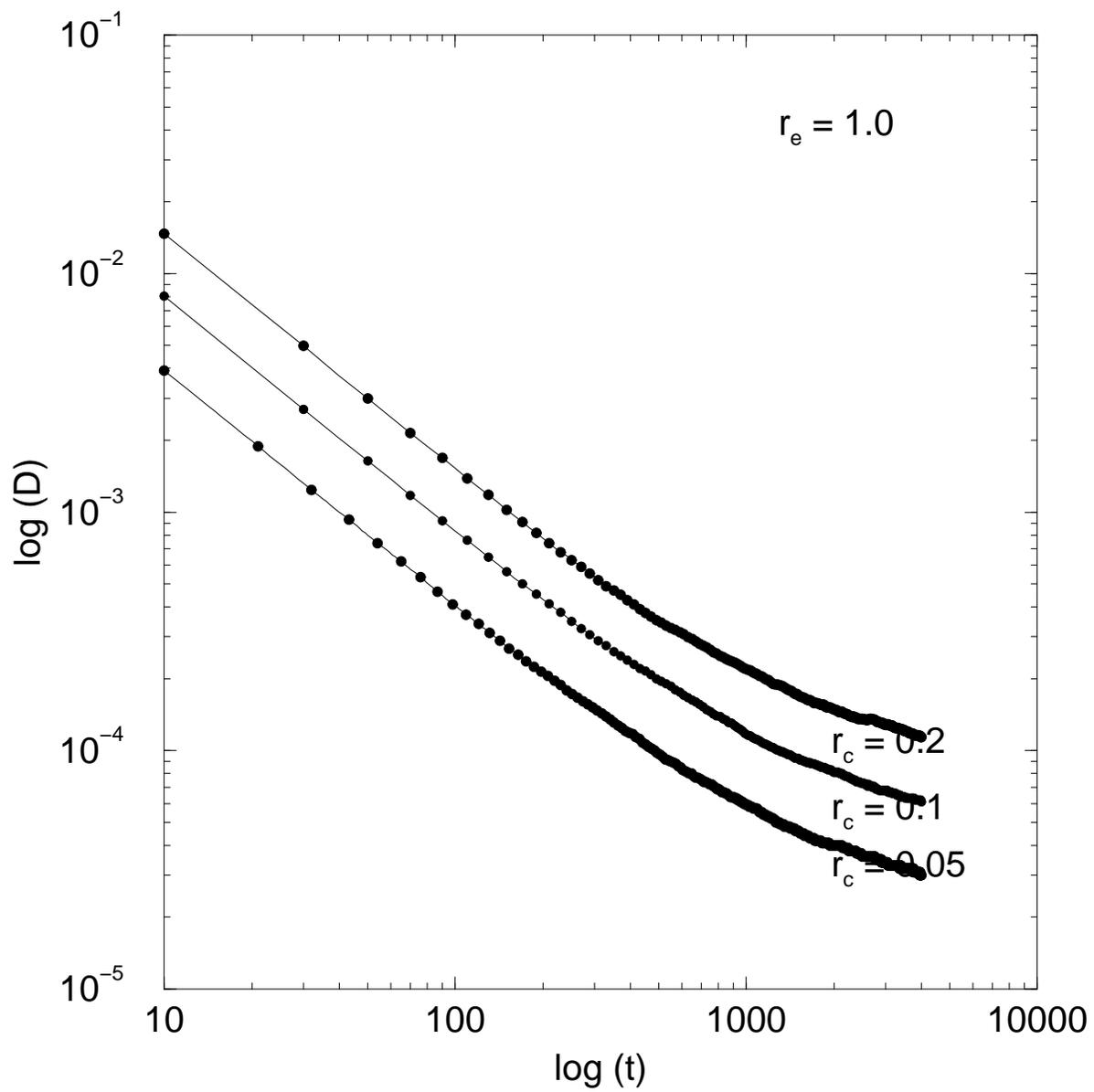
\begin{figure}
\caption{Log-log plot of the diffusion coefficient vs time 
(at early evolution times) for $r_e = 1.0$ 
and different values of $r_c$. 
The straight decay indicates an scaling behavior 
$D (t) \sim t^{-\beta}$, with $\beta = 0.812 \pm 0.021$.} 
\end{figure}



\begin{references}

\bibitem{SOL2} L. Kuipers, M.S. Hoogeman and J.W. Frenken, Phys. Rev. Lett.
{\bf 71}, 3517 (1993).

\bibitem{SOL3} M. Giesen-Seibert, R. Jentjens, M. Poensgen, and H. Ibach,
Phys. Rev. Lett. {\bf 71}, 3521 (1993).

\bibitem{BIN} K. Binder and M.H. Kalos, J. Stat. Phys {\bf 22}, 263 (1980).

\bibitem{SOL4} H.J. Mamin, p.H. Guethner, and D. Rugar, Phys. Rev. Lett.
{\bf 65}, 2418 (1990).

\bibitem{SOL5} J.I. Pascual {\it et al}, Phys. Rev. Lett. {\bf 71}, 1852 (1993).

\bibitem{SOL6} R.C. Jaklevic and L. Elie, Phys. Rev. Lett. {\bf 60}, 120 (1988).

\bibitem{EXP} J.M. Wen, S.L. Chang J.W. Burnett, J.W. Evans and P.A. Thiel, 
Phys. Rev. Lett. {\bf 73}, 2591 (1994).

\bibitem{WANG6} S.C. Wang and G. Ehrlich, Surf. Sci. {\bf 301}, 239 (1985).

\bibitem{FINK5} H.W. Fink and  G. Ehrlich, Surf. Sci. {\bf 150}, 419 (1990).

\bibitem{KANG14} H.C. Kang, P.A. Thiel and J.W. Evans, J. Chem. Phys. {\bf 93}, 9018
(1990).

\bibitem{VOTER18} A.F. Voter, Phys. Lett. B {\bf 34}, 6819 (1986).

\bibitem{SQR} The square lattice is an appropriate model for the Ag(100) surface.

\bibitem{SOL1} J.M. Soler, Phys. Rev. B {\bf 50}, 5578 (1994). 

\bibitem{SOL2} J.M. Soler, Phys. Rev. B {\bf 53}, R10540 (1996).

\bibitem{VANSIC} C.DeW. Van Siclen, Phys. Rev. Lett. {\bf 75}, 1574 (1995).

\bibitem{AVR} The average has been taken over the three $r_c / r_e$ relations used.

\end{references}
\end{document}